\documentclass[aps,prb,reprint,showpacs]{revtex4-1} 
\usepackage[utf8]{inputenc}
\usepackage{graphicx}
\usepackage{amsmath}
\usepackage{amssymb}
\usepackage{verbatim}
\usepackage[]{hyperref}
\usepackage{hhline}
\usepackage{subcaption}
\usepackage{braket}
\usepackage{array}
\usepackage{adjustbox}
\usepackage{tabularx}
\newcolumntype{L}[1]{>{\raggedright\let\newline\\\arraybackslash\hspace{0pt}}m{#1}}
\newcolumntype{C}[1]{>{\centering\let\newline\\\arraybackslash\hspace{0pt}}m{#1}}
\newcolumntype{R}[1]{>{\raggedleft\let\newline\\\arraybackslash\hspace{0pt}}m{#1}}

\begin{document}

\begin{abstract}
We present an extensive set of surface and chemisorption energies calculated using state of the art many-body perturbation theory. In the first part of the paper we consider ten surface reactions in the low coverage regime where experimental data is available. Here the random phase approximation (RPA) is found to yield high accuracy for both adsorption and surface energies. In contrast all the considered density functionals fail to describe both quantities accurately. This establishes the RPA as a universally accurate method for surface science. In the second part, we use the RPA to construct a database of 200 high quality adsorption energies for reactions involving OH, CH, NO, CO, N$_2$, N, O and H over a wide range of 3d, 4d and 5d transition metals. Due to the significant computational demand, these results are obtained in the high coverage regime where adsorbate-adsorbate interactions can be significant. RPA is compared to the more advanced renormalised adiabatic LDA (rALDA) method for a subset of the reactions and they are found to describe the adsorbate-metal bond as well as adsorbate-adsorbate interactions similarly. The RPA results are compared to a range of standard density functional theory methods typically employed for surface reactions representing the various rungs on Jacob's ladder. The deviations are found to be highly functional, surface and reaction dependent. 
Our work establishes the RPA and rALDA methods as universally accurate full ab-initio methods for surface science where accurate experimental data is scarce. The database is freely available via the Computational Materials Repository (CMR).
\end{abstract}

\title{Benchmark Database of Transition Metal Surface and Adsorption Energies from Many-Body Perturbation Theory}
\author{Per S. Schmidt}
\author{Kristian S. Thygesen}
\email{psisc@fysik.dtu.dk}
\affiliation{Center for Atomic-scale Materials Design (CAMD), Technical University of Denmark, DK-2800 Kongens Lyngby, Denmark}

\date{\today}

\maketitle

\section{Introduction}
The application of density functional theory (DFT) to problems in surface science is ever increasing. In particular, the ability to predict stability and reactivity of transition metal surfaces is an important and fundamental problem in many areas, not least heterogeneous catalysis. 
Immense efforts have gone into the development of better exchange-correlation (xc)-functionals and today hundreds of different types exist, with the most popular types being the generalized gradient approximations (GGAs), the meta GGAs, (screened) hybrids, GGA+U, and the non-local van der Waals density functionals. With a few exceptions, they all contain parameters that have been optimized for a particular type of problem or class of material. Moreover they rely on fortuitous and poorly understood error cancellation effects. This limits the generality and predictive power of the standard xc-functionals whose performance can be highly system dependent.

Recently, the random phase approximation (RPA) has been advanced as a total energy method that goes beyond standard DFT\cite{furche,gonze}. Within the RPA, the correlation energy is obtained from the linear density response function while exchange is treated exactly. It is computationally much more demanding than conventional DFT (including orbital dependent functionals), but significantly cheaper than wave function-based  quantum chemistry methods. The RPA is presently considered the gold standard for solid state systems due to its ab-initio nature, good description of static correlation and excellent account of long range dispersive forces\cite{Ren2012,marini}. Compared to standard DFT, the RPA reduces self-interaction errors due to its exact treatment of exchange and does as such not rely on error cancellation between exchange and correlation. Additionally, it incorporates dynamical screening and accounts for long-range correlations such as van der Waals interactions through its non-locality. \\
For molecular systems, advanced quantum-chemical techniques, such as coupled-cluster theory including single, double and perturbative triple particle-hole excitation operators, CCSD(T), has been applied with great success. There does however only exist very few cases where this method has been applied to solids or surfaces due to its high computational complexity and polynomial scaling with system size making it extremely expensive\cite{gruneisnature}.

The RPA has been found to be most accurate for energy differences between isoelectronic systems, i.e. systems with similar electronic structure. Thus structural parameters are generally accurately described as is bonds of dispersive or mixed dispersive-covalent nature. In contrast, strong covalent bond energies are typically underestimated by RPA, implying that atomisation energies of covalently bonded crystals and molecules are systematically underestimated. This deficiency is related to the relatively poor description of the short range correlation hole by the RPA \cite{hedin, singwi}. The introduction of an xc-kernel, such as the simple renormalised adiabatic local density approximation (rALDA), in the density response function, greatly improves the short range correlation hole and leads to a significant reduction of the error on covalent bond energies \cite{tols1,tols2,tols3}. 

While most previous RPA total energy studies have focused on isolated molecules and bulk solids, there have been some RPA reports on surface and adsorption problems\cite{kresse_ads,Ren2009,tolsrpa, rohlfing, ma_rpa, kim_rpa}. Graphene adsorption on metal surfaces is considered notoriously difficult due to the mixed dispersive-covalent nature of the graphene-metal bond. Nevertheless, the predicted RPA binding distances are in excellent agreement with available experimental data \cite{tolsrpa}.
The RPA has successfully resolved the “CO adsorption puzzle”: in contrast to most DFT functionals RPA predicts the correct binding site of CO on Pt(111) and Cu(111). For the case of CO on Pt(111) and Pd(111), Schimka et al.\cite{kresse_ads} demonstrated how semi-local xc-functionals underestimate the surface energy relative to experiments and at the same time overestimate the adsorption energy. By modifying the xc-functional, either the predicted adsorption energies or surface energies can be improved but always at the expense of the other\cite{kresse_ads}. In contrast, the RPA improves the description of both properties simultaneously. 

\begin{figure}
\centering
\includegraphics[width=1.0\columnwidth]{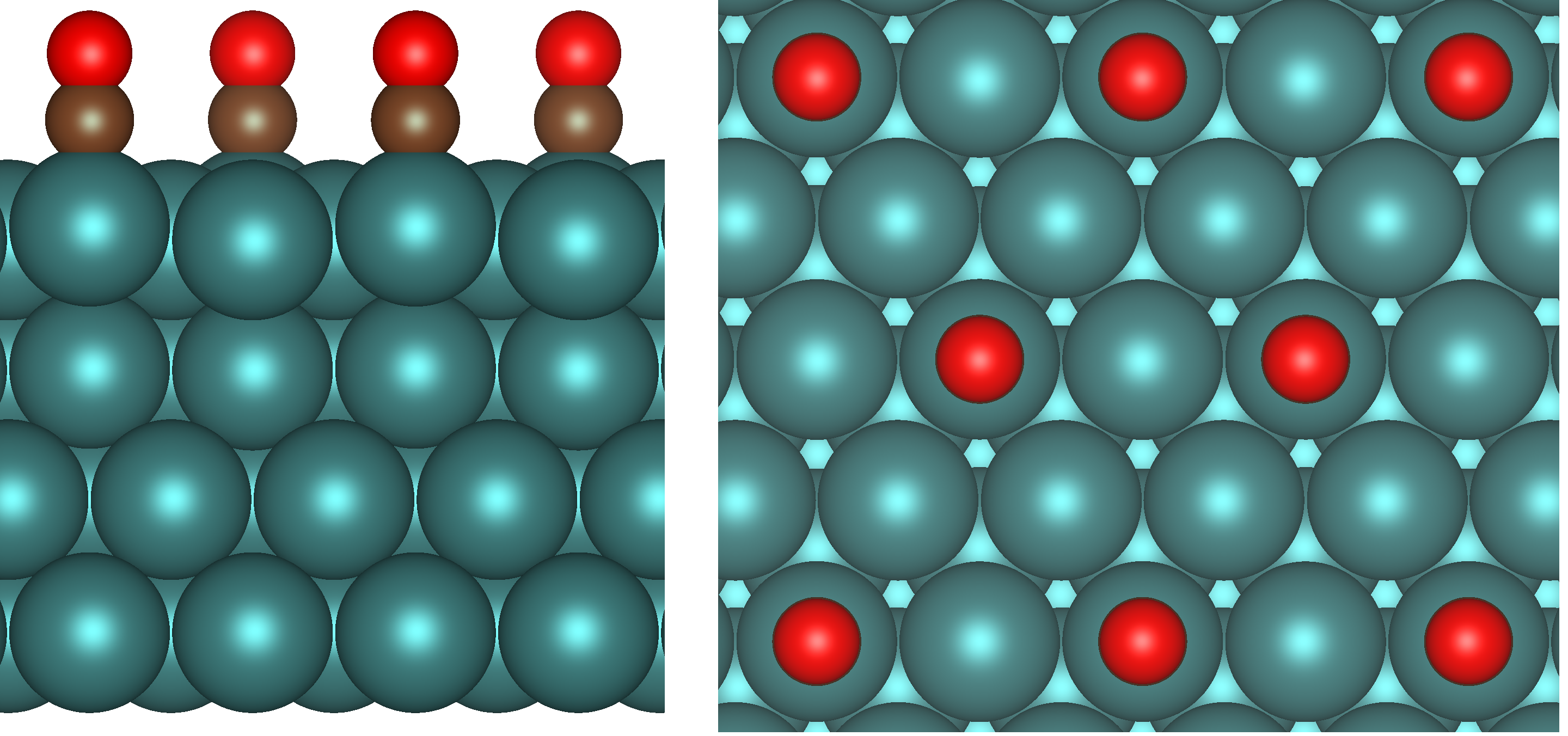}
\caption{\label{fig:structures}Side- and topview of CO adsorbed in the fcc site of a (111) surface with a coverage of 1/4.}
\end{figure}

In this work we present an extensive and systematic study of surface adsorption based on the RPA and rALDA methods with comparisons to experiments and selected xc-functionals. The paper consists of two parts. In the first part, we apply the RPA to a range of adsorption reactions at 1/4 coverage, as illustrated in Fig. \ref{fig:structures}, where experimental results are available. On average, the RPA predicts adsorption energies that deviate by 0.2 eV from experiments. The same average accuracy is achieved by the RPBE and BEEF-vdW xc-functionals. While the two xc-functionals perform rather similarly, there are significant and non-systematic deviations between the xc-functionals and RPA for the individual reactions indicating the different nature of the RPA compared to the semi-local and vdW density functionals. The good performance of the RPBE and BEEF-vdW for adsorption energies does not carry over to surface energies, which are hugely underestimated. On the other hand RPA remains accurate also for surface energies. 

Having established the reliability of the RPA for surface reactions, the second part of the paper is devoted to the development of a comprehensive reference database of adsorption bond energies to transition metal surfaces. The motivation for this endeavour is manifold. 
First, the database will make it easier for code developers to compare and benchmark their results, which is important in order to enhance the reproducibility of RPA calculations. Important progress along these lines have already started for standard DFT calculations\cite{Lejaeghereaad}. The much higher complexity of many-body based methods compared to DFT calculations makes it even more important to facilitate such developments for RPA calculations. Secondly, the development of better xc-functionals relies crucially on access to large, well defined and consistent datasets. Using experimental data is not ideal because they are influenced by factors not considered in the calculations and thus there is a risk of obtaining the right result for the wrong reason. For methods targeting surface science problems, the situation is even worse because of the scarcity of accurate experimental data for adsorption and surface energies not to mention transition state energies. 
It is therefore critical to develop theoretical reference datasets for unique reactions and surface structures calculated with the most accurate computational methods available. The concept of theoretical reference databases is widely used in quantum chemistry, but is presently lacking in materials and surface science. 

The obvious challenge related to the establishment of RPA (or beyond-RPA) reference datasets for surface science is the large computational cost of such calculations. We overcome this problem by focussing on the high coverage limit where the small size unit cells renders the problem tractable while still permitting an assessment of the critical metal-adsorbate and adsorbate-adsorbate interactions. Specifically, we calculate a total of 200 different full coverage adsorption reactions involving OH, CH, NO, CO, N$_2$, N, O and H adsorbed on a wide range of 3d, 4d and 5d transition metals.  We compare the RPA and rALDA results to a wide set of xc-functionals implemented in the electronic structure code GPAW representing the different rungs of Jacob's ladder (LDA, PBE\cite{PBE}, RPBE\cite{RPBE}, vdW-DF2\cite{vdWDF2}, mBEEF\cite{mBEEF}, BEEF-vdW\cite{BEEF} and mBEEF-vdW\cite{mBEEFvdW}).

\section{Part I: Experimentally relevant reactions}
\subsection{Methods}
In the first part of this work we use the RPA method to calculate adsorption and surface energies of systems where experimental data is available for comparison\cite{jess}. The surface is represented by a slab containing four atomic layers and the experimentally relevant coverage and adsorption site is used. The reactions considered are CO, NO, O and H on (111)-surfaces of Pt, Rh, Ir, Cu, Pd and Ni. See the section on computational details for more information. 

The adsorption energy is defined with reference to the corresponding molecule in its gas phase, 
$$ E_{\text{ads}} = E_{\text{adsorbate@slab}} - (E_{\text{slab}} + E_{\text{adsorbate(g)}})$$
while the surface energy is defined as 
\begin{equation*}
E_{\text{surf}} =  \frac12 \bigg( E_{\text{slab}} - N_{\text{layers}} E_{\text{bulk}}\bigg)  \label{eq:surf}
\end{equation*}
\subsection{Results and discussion}
\subsubsection{Adsorption energies}
\begin{table*}
\begin{tabular}{L{2.7cm}L{0.5cm}L{2.4cm}C{0.7cm}C{1.0cm}C{1.2cm}C{1.5cm}C{1.0cm}C{3.6cm}}
\hline\hline \noalign{\smallskip}
 & & & site & PBE & RPBE &BEEF-vdW & RPA & Exp.\\
\hline\noalign{\smallskip}
CO + Pt(111) & $\rightarrow$ & CO/Pt(111) & top & -1.68 & -1.29 & -1.20 & -1.36 & -1.20, -1.22, -1.26, -1.28\\
CO + Rh(111) & $\rightarrow$ & CO/Rh(111) & top & -1.92 & -1.56 & -1.50 & -1.40 & -1.29, -1.33, -1.60\\
CO + Ir(111) & $\rightarrow$ & CO/Ir(111) & top & -2.11 & -1.74 & -1.72 & -1.48 &-1.61\\
CO + Cu(111) & $\rightarrow$ & CO/Cu(111) & top & -0.74 & -0.41 & -0.45 & -0.25 &-0.51, -0.57\\
CO + Pd(111) & $\rightarrow$ & CO/Pd(111) & fcc & -1.91 & -1.48 & -1.49 & -1.58 &-1.34, -1.38, -1.41, -1.43, -1.43, -1.46, -1.47, -1.57 \\
\hline\noalign{\smallskip}
NO + Pd(111) & $\rightarrow$ & NO/Pd(111) & fcc & -2.17 & -1.75 & -1.75 & -1.99 & -1.81\\
NO + Pt(111) & $\rightarrow$ & NO/Pt(111) & fcc & -1.88 & -1.45 & -1.43 & -1.33 & -1.16\\
\hline\noalign{\smallskip}
$\frac12 \mathrm{O}_2$ + Ni(111) & $\rightarrow$ & O/Ni(111) & fcc & -2.16 & -1.81 & -1.95 & -2.50 & -2.49\\
\hline\noalign{\smallskip}
$\frac12 \mathrm{H}_2$ + Ni(111) & $\rightarrow$ & H/Ni(111) & fcc & -0.53 & -0.34 & -0.27 & -0.36 &  -0.45, -0.48\\
$\frac12 \mathrm{H}_2$ + Pt(111) & $\rightarrow$ & H/Pt(111) & top & -0.46 & -0.28 & -0.14  & -0.51 & -0.36, -0.37\\
\hline\noalign{\smallskip}
MAE & & & & \phantom{-}0.44 & \phantom{-}0.23 & \phantom{-}0.19 & \phantom{-}0.21 & -\\
MSE & & & &-0.38 & -0.04 & -0.01 & -0.10 &-\\
\hline\hline
\end{tabular}
\caption{\label{tab:adsextrap}Ten adsorption energies calculated at the experimentally relevant coverage to allow for a direct comparison using three different DFT xc functionals (PBE, RPBE and BEEF-vdW) and the RPA. The experimental values are from Ref.\cite{jess} and the references therein. A finite temperature correction have been added to the experimental values. All values are in eV. }
\end{table*}

\begin{figure*}
 \centering
 \includegraphics[width=2.1\columnwidth]{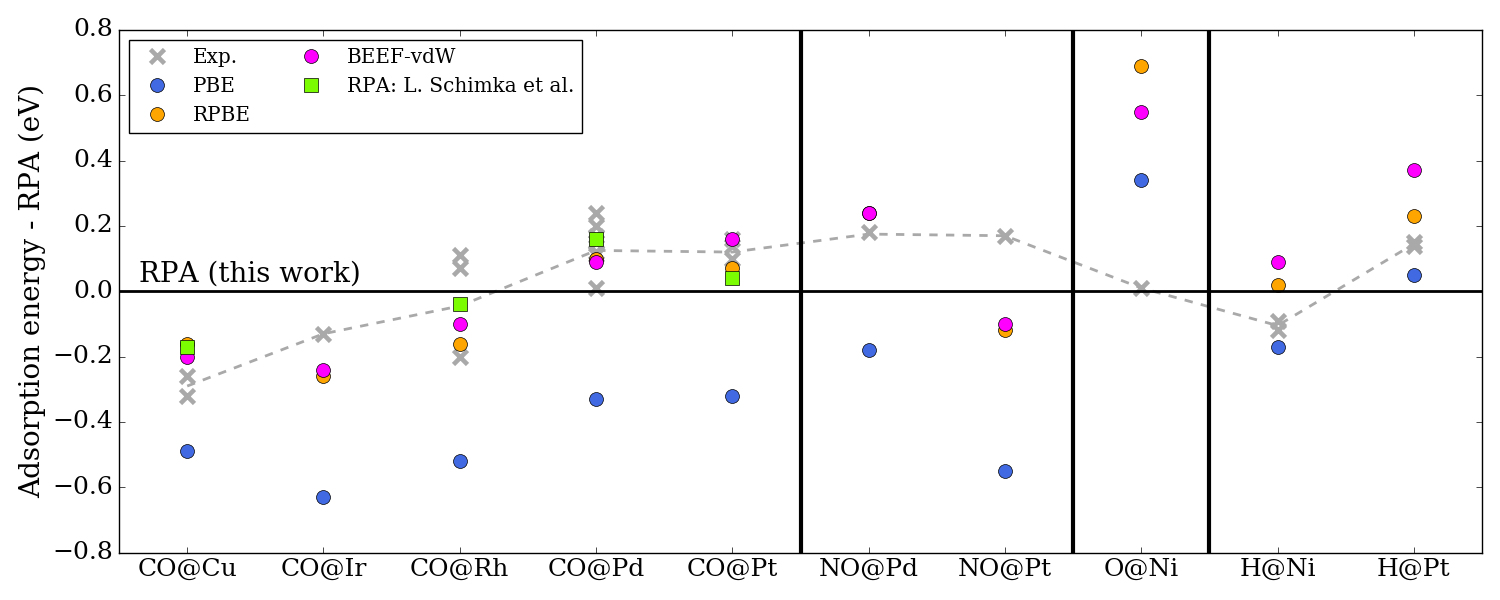}
 \caption{\label{fig:ads_2x2_all} Deviations in adsorption energies from the RPA results of this work. The green squares are RPA results from \cite{kresse_ads}.}
\end{figure*}

Adsorption energies from PBE, RPBE, BEEF-vdW and RPA and are shown in Table \ref{tab:adsextrap} along with selected experimental values. The experimental reference data are mainly from equilibrium adsorption studies, temperature programmed desorption (TPD) and single crystal adsorption calorimetry (SCAC) (see Ref. \cite{jess} and references therein). When comparing to theoretical adsorption energies, it should be kept in mind that, even when carefully executed, such experiments are always subject to inherent uncertainties stemming from variations in the surface crystal structure, presence of surface defects or impurities, different binding sites, side reactions, etc. Consequently, experimental adsorption energies can vary by up to 0.3 eV for the same reaction although for most reactions the variation is around 0.1 eV. The mean absolute error (MAE) and mean signed error (MSE) of the theoretical results have been calculated relative to the average experimental adsorption energy reported for each reaction.  

Fig. \ref{fig:ads_2x2_all} shows the deviations of the calculated and experimental adsorption energies from the RPA values of this work. 
The RPA, RPBE, and BEEF-vdW all have a MAE of around 0.20 eV which is significantly lower than the 0.44 eV obtained with PBE. The systematic overbinding by PBE is a well known problem. The case of O on Ni(111) deviates from the general trends which is related to the rather poor description of the O$_2$ reference by the DFT functionals. It should be noted that the BEEF-vdW has been fitted to a dataset containing the first 8 adsorption reactions considered. 

The performance of the RPA is actually quite remarkable in view of its pure ab-initio nature. With a MSE of -0.10 eV, it appears that the RPA has a weak tendency to overbind the adsorbates. However, given the small size of the dataset and the inherent uncertainties in experimental data one should be careful to draw too strict conclusions. The agreement with previously published RPA values (for CO on Cu, Rh, Pd, Pt) is acceptable with a MAE of 0.10 eV and MSE of 0.002 eV. 
These deviations could stem from differences in the applied geometries (although PBE relaxed structures were used in both studies) and the fact that the RPA results from Ref. \cite{kresse_ads} were obtained by extrapolating RPA energies for a $\sqrt3\times \sqrt3$ unit cell to a $2\times 2$ cell using PBE data. 
\subsubsection{Surface energies}
\begin{figure}[h!]
\includegraphics[width=\columnwidth]{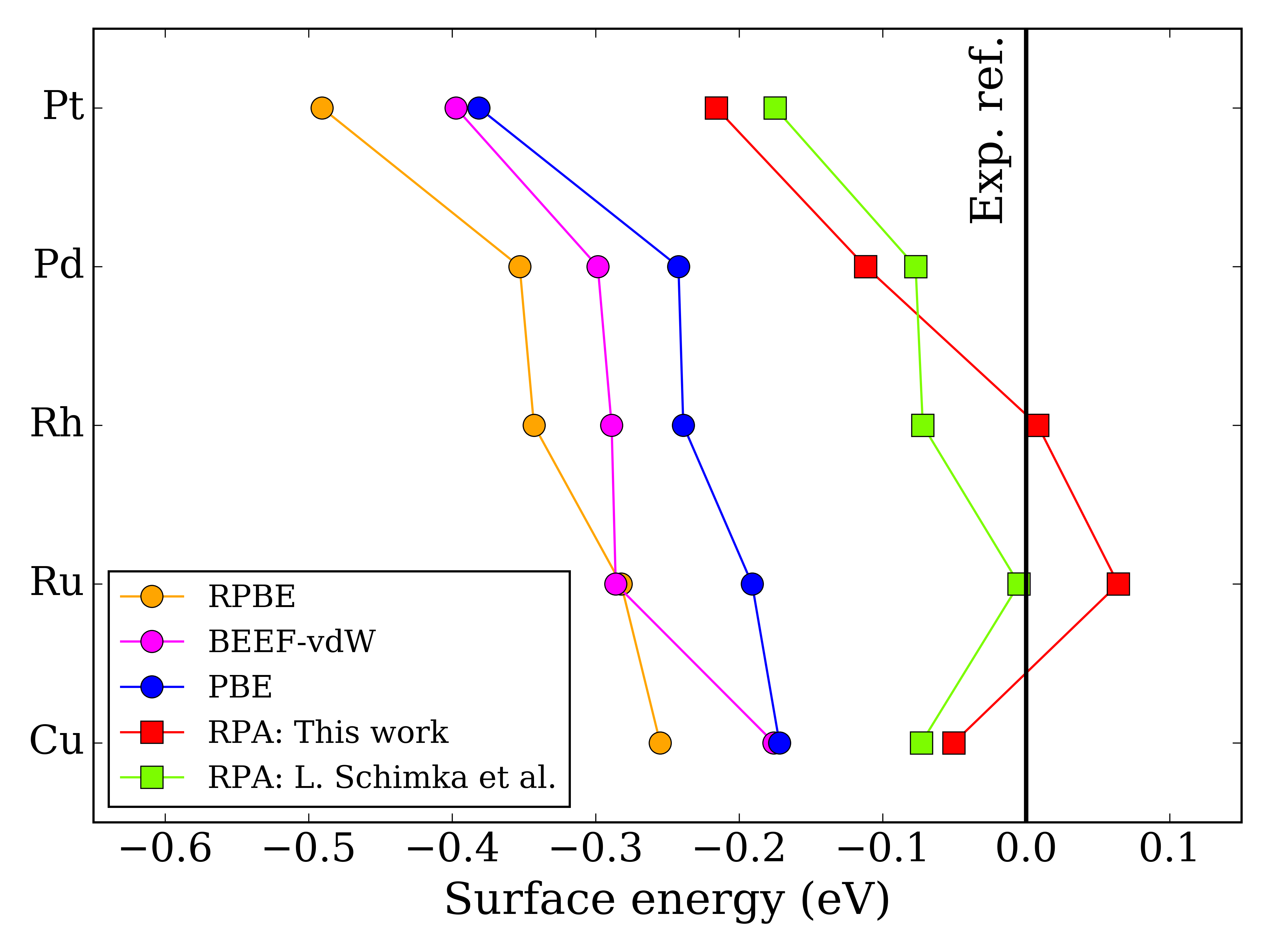}
\caption{\label{fig:kresse_compare} Surface energies of five different fcc(111) surfaces calculated with four different methods and compared to experimental values and RPA values from \cite{kresse_ads}.}
\end{figure}
Surface energies of some selected (111) surfaces were calculated with RPA, rALDA and different DFT xc-functionals. Fig. \ref{fig:kresse_compare} shows the surface energies obtained in the present work together with the RPA results of Ref. \cite{kresse_ads}, plotted relative to the experimental values. Our RPA results show excellent agreement with the previous RPA values and confirm that RPA predicts surface energies in much better agreement with experiments than all the considered DFT functionals. Inclusion of the rALDA kernel does not have a large effect on the surface energies which are very similar to the RPA values (not shown). 

As previously observed, it is striking that the xc-functionals which perform better for adsorption energies (RPBE and BEEF-vdW) perform worse for surface energies and vice versa. This circumstance highlights the limitations of presently employed functionals and underlines the advantage of more advanced methods, like the RPA and the rALDA. The latter exhibit a greater degree of universality in the sense that they offer high (though not perfect) accuracy across a broad range of systems and bonding types.   
 
\subsubsection{Coverage effects}\label{sec:coverage}
\begin{figure}
\includegraphics[width=\columnwidth]{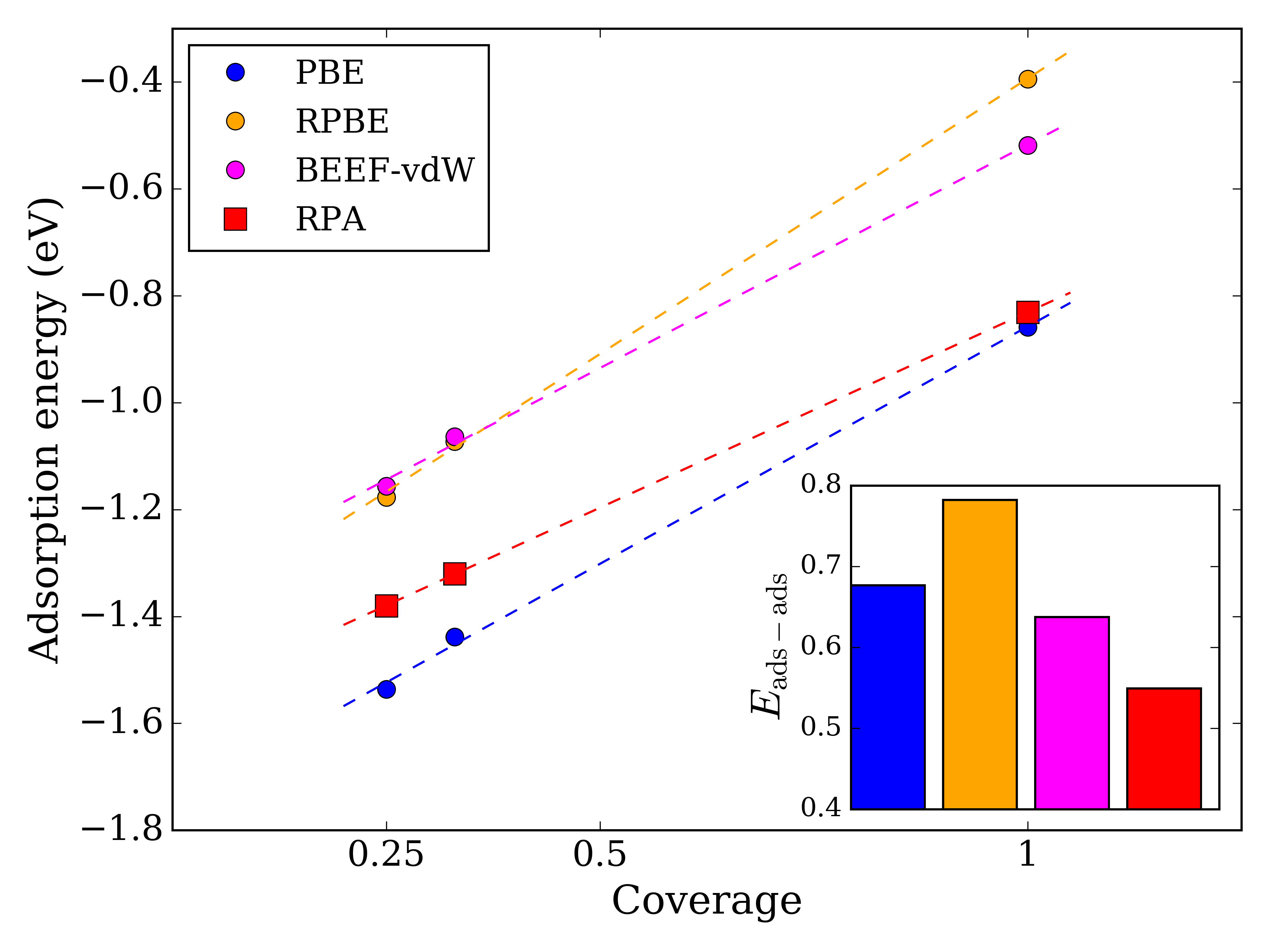}
\caption{\label{fig:025_to_1}Adsorption energies of CO on the topsite of Pt(111) with a coverage ranging from 1/4 to 1. The magnitudes of the adsorbate-adsorbate interactions are shown in the inset, calculated as the difference between the adsorption energies at full and 1/4 coverage. }
\end{figure}

As mentioned in the introduction, the computational cost of the RPA and rALDA makes large scale application of these methods a daunting task. As a consequence, for benchmarking purposes it is desirable to explore minimalistic models of adsorption systems which still capture the essential physical mechanisms governing surface reactions, i.e. the formation and breaking of bonds between an extended metal surface and the adsorbate. In Fig. \ref{fig:025_to_1} we show the coverage dependence of the adsorption energy for the case of CO on top of a Pt(111) surface.  Clearly, there is a significant dependence of the binding energy on the coverage due to repulsive adsorbate-adsorbate interactions at higher coverages. However, the relative ordering of the adsorption energy obtained with different methods is essentially unchanged. This shows that adsorption in the high coverage limit involves much the same physics as the low coverage regime. Having made this point, it should be stressed that the high coverage regime is of interest in itself, as high coverage configurations become relevant at high gas pressure conditions. Moreover, adsorbate-adsorbate interactions, which contribute to the adsorption energy at high coverage, are generally important to incorporate for a correct description of reaction kinetics. Finally, high coverage configurations do, at least to some extent, resemble the stretched bond configurations of dissociative transition states. Thus one could expect a more accurate description of high coverage configurations correlates with more accurate descriptions of transition states and barrier heights. In this context, it is interesting to note from    
Fig. \ref{fig:025_to_1} that the magnitude of the repulsive adsorbate-adsorbate interactions is largest with the pure GGAs (PBE and RPBE), smaller with the van der Waals density functional (BEEF-vdW), and smallest with the RPA. We believe this is due to the stabilising effect of attractive van der Waals interactions between the adsorbates at higher coverage. 

\clearpage
\section*{Part II: Database of adsorption energies}
\subsection{Methods}
In this section we present 200 adsorption energy calculations of full coverage reactions in order to compare the performance of different computational methods. The adsorption is always on the top site of the fcc(111) surface represented by three layers using the bulk PBE lattice constant. The position of the top surface layer and the adsorbate has been relaxed using the PBE. Adsorption energies have been calculated with the RPA and rALDA methods as well as the xc-functionals: LDA, PBE, RPBE, vdW-DF2, mBEEF, BEEF-vdW and mBEEF-vdW. RPA and rALDA show close agreement so to limit the presentation emphasis will be on RPA and its comparison to PBE, RPBE and BEEF-vdW. See the Computational Methods section for a description of the many-body methods and further computational details.

The adsorption energies were calculated for the following reactions on transition metal surfaces with the reference molecule in their gas phase:\\

\begin{tabular}{lrl}
(1)\phantom{\quad} & H$_2$O + slab &$\rightarrow$ OH/slab + $\frac{1}{2}$ H$_2$\\
(2) & CH$_4$ + slab &$\rightarrow$ CH/slab + $\frac{3}{2}$ H$_2$\\
(3) & NO + slab &$\rightarrow$ NO/slab\\
(4) & CO + slab &$\rightarrow$ CO/slab\\
(5) & N$_2$ + slab &$\rightarrow$ N$_2$/slab\\
(6) & $\frac{1}{2}$N$_2$ + slab &$\rightarrow$ N/slab\\
(7) & $\frac{1}{2}$O$_2$ + slab &$\rightarrow$ O/slab\\
(8) & $\frac{1}{2}$H$_2$ + slab &$\rightarrow$ H/slab\\
 &
\end{tabular}

\begin{figure}[t!]
\centering
\includegraphics[width=1.0\columnwidth]{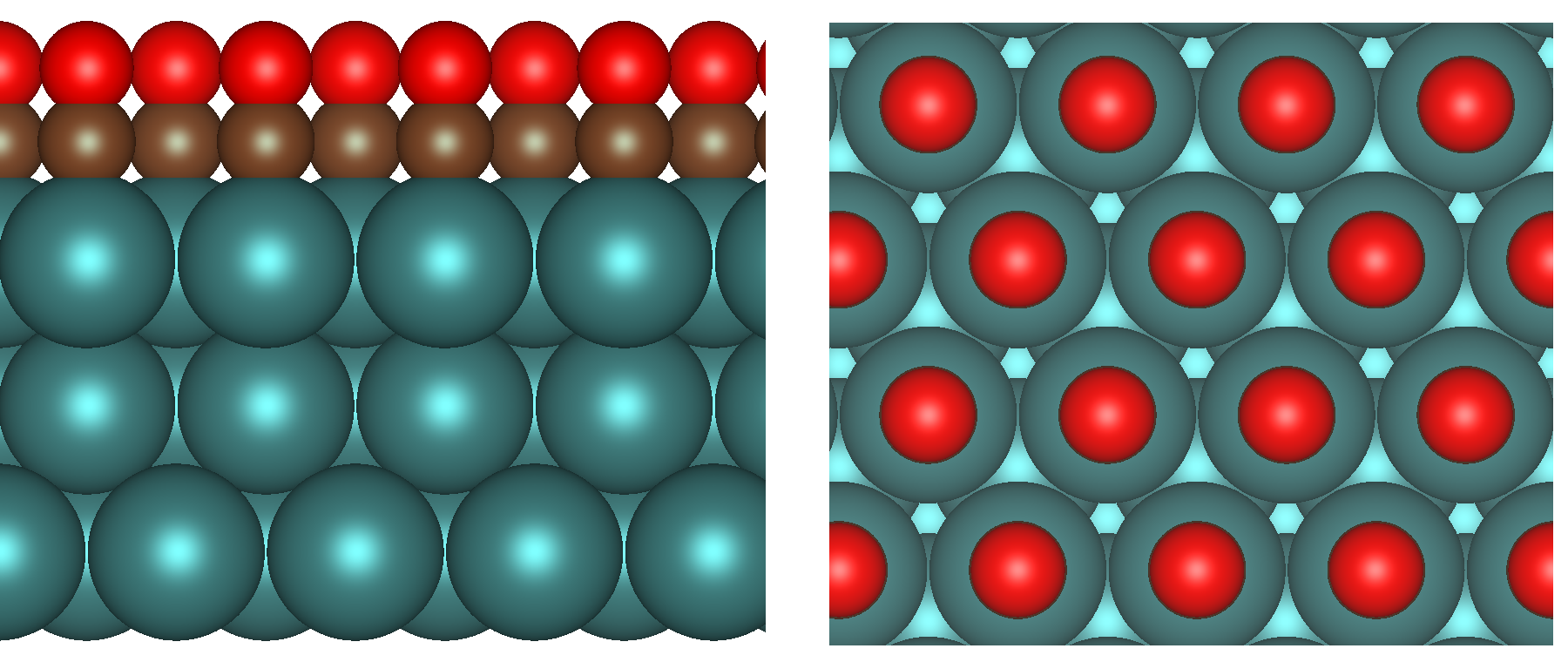}
\caption{\label{fig:fullcov_structures}Side- and topview of CO adsorbed in the top site at full coverage.}
\end{figure}

An example of the full coverage adsorption geometry is shown in Fig. \ref{fig:fullcov_structures}. The adsorbate-adsorbate distance is around 3 Å, and consequently adsorbate-adsorbate interactions will certainly affect the adsorption energies (see discussion in the section on coverage effects). 
In Table \ref{tab:RPA_rALDA} we report the difference in adsorption energy calculated with RPA and rALDA for all the different adsorbates on Mn(111) and a few different surfaces for CO adsorption. The rALDA results are in excellent agreement with RPA, with the largest deviation seen for OH adsorption (0.11 eV). This agrees well with previous results for CO adsorption on Pt(111) and graphene on Ni(111)\cite{tols1}. The fact that the two methods agree suggests that the relevant physics is already described by the RPA, i.e. that local correlations play a smaller role for molecule-metal bonding. However, it is interesting to note that RPA shows a small but systematic tendency to overbind the adsorbates relative rALDA. This is in stark contrast to the situation for covalently bonded atoms and semi-conductors, where RPA underestimates the bond strength by 0.44 eV/atom and 0.30 eV/atom in average, relative to rALDA, respectively. For metals, the cohesive energy is also underestimated by RPA, but by somewhat smaller degree (MSE of -0.15 eV relative to rALDA), see Table \ref{tab:MAEMSE}. We propose the following explanation for the opposed trend for molecule-metal bonding: For reactions involving the breaking of chemical bond in the initial adsorbate molecule, e.g. for reactions of the type $\frac12 \mathrm{A}_2$ $\to$ A/metal (reactions 1,2,6,7,8) RPA will overestimate the reaction energy because the A-A bond strength is underestimated more than the A-metal bond. For pure adsorption reactions of the form A$_2$ $\to$ A$_2$/metal (reactions 3,4,5), RPA will overestimate the adsorption energy because the reduction of the internal A-A bond upon adsorption is underestimated more than the A$_2$/metal bond. In both cases the reason for the (slight) overestimation of the reaction energy can thus be traced to a larger underestimation by the RPA in describing pure covalent bonds compared to bonds with partial metallic character.
With this in mind, we focus on the RPA in the rest of the paper.

\begin{table*}
\begin{tabular}{llc|cccccccc}
\hline\hline \noalign{\smallskip}
 Ads. & Surf. & rALDA & RPA & LDA & PBE & RPBE & vdW-DF2 & BEEF-vdW & mBEEF & mBEEF-vdW\\
\hline\noalign{\smallskip}
H & Mn & 0.64 & 0.64 & 0.09 & 0.41 & 0.55 & 0.58 & 0.56 & 0.44 & 0.30 \\
O & Mn & -0.76 & -0.81 & -1.84 & -1.05 & -0.70 & -1.04 & -0.87 & -1.00 & -1.16 \\
N & Mn & 2.30 & 2.30 & 1.06 & 1.82 & 2.15 & 2.08 & 2.03 & 1.93 & 1.74 \\
N$_2$ & Mn & 0.75 & 0.74 & -0.63 & 0.50 & 1.04 & 0.83 & 0.70 & 0.55 & 0.10 \\
CO & Mn & -0.14 & -0.21 & -1.61 & -0.50 & 0.03 & 0.06 & -0.22 & -0.50 & -0.95 \\
NO & Mn & -0.96 & -0.99 & -2.87 & -1.50 & -0.96 & -1.06 & -1.19 & -1.33 & -1.68 \\
CH & Mn & 3.62 & 3.59 & 3.22 & 3.56 & 3.72 & 3.56 & 3.49 & 3.71 & 3.51 \\
OH & Mn & 1.36 & 1.24 & 0.46 & 1.12 & 1.44 & 0.98 & 1.15 & 1.14 & 0.95 \\
CO & Sc & -0.60 & -0.61 & -1.48 & -0.97 & -0.71 & -0.94 & -0.89 & -0.94 & -1.10 \\
CO & Ti & -0.77 & -0.81 & -1.68 & -1.00 & -0.66 & -0.80 & -0.90 & -1.04 & -1.32 \\
CO & V & -0.79 & -0.85 & -1.87 & -1.01 & -0.58 & -0.68 & -0.84 & -1.03 & -1.39 \\
CO & Cr & -0.34 & -0.38 & -1.75 & -0.74 & -0.25 & -0.27 & -0.50 & -0.77 & -1.17 \\
\hline\noalign{\smallskip}
\multicolumn{3}{l}{MAE (eV) vs. rALDA} & 0.04 & 1.10 & 0.31 & 0.12 & 0.16 & 0.15 & 0.28 & 0.54 \\
\multicolumn{3}{l}{MAE (\%) vs. rALDA} & 8 & 218 & 58 & 24 & 29 & 22 & 56 & 115 \\
\hline\hline
\end{tabular}
\caption{\label{tab:RPA_rALDA}Adsorption energies (in eV) for a few reactions at full coverage calculated with rALDA, RPA and 7 different DFT xc functionals.}
\end{table*}

\begin{table}
\begin{tabularx}{\columnwidth}{L{2.37cm}llll}
& \multicolumn{2}{l}{$E^{\mathrm{RPA}} - E^{\mathrm{rALDA}}$} & \multicolumn{2}{l}{$E^{\mathrm{RPA}} - E^{\mathrm{Exp.}}$} \\
\hline\hline \noalign{\smallskip}
 & MAE & MSE & MAE & MSE \\
\hline\noalign{\smallskip}
{\small Molecules} & 0.48 & -0.44 & 0.52 & -0.52 \\
{\small Gapped solids} & 0.30 & -0.30 & 0.43 & -0.43 \\
{\small Metals} & 0.18 & -0.15 & 0.24 & -0.24 \\
\hline\hline
\end{tabularx}
\caption{\label{tab:MAEMSE} Difference in atomization and cohesive energies between RPA, rALDA and experiments for three different types of materials. Data taken from \citep{tols1}.}
\end{table}

\subsection{Results and discussion}
\subsubsection{Adsorption energies}
\begin{figure*}
\centering
\begin{subfigure}{2.1\columnwidth}
\includegraphics[clip, trim=1.8cm 0.cm 3.5cm 0.cm, width=0.9\columnwidth]{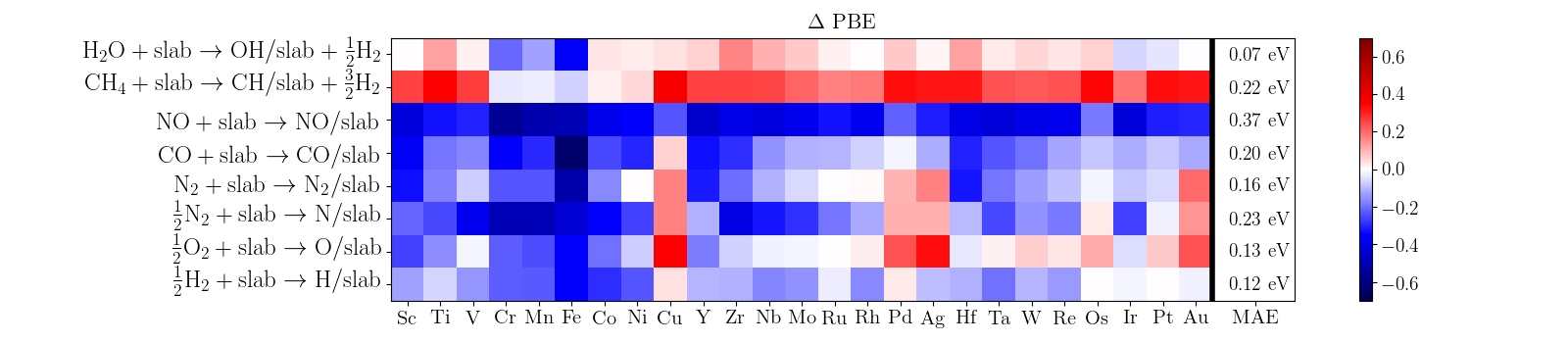}
\end{subfigure}
\begin{subfigure}{2.1\columnwidth}
\includegraphics[clip, trim=1.8cm 0.cm 3.5cm 0.cm, width=0.9\columnwidth]{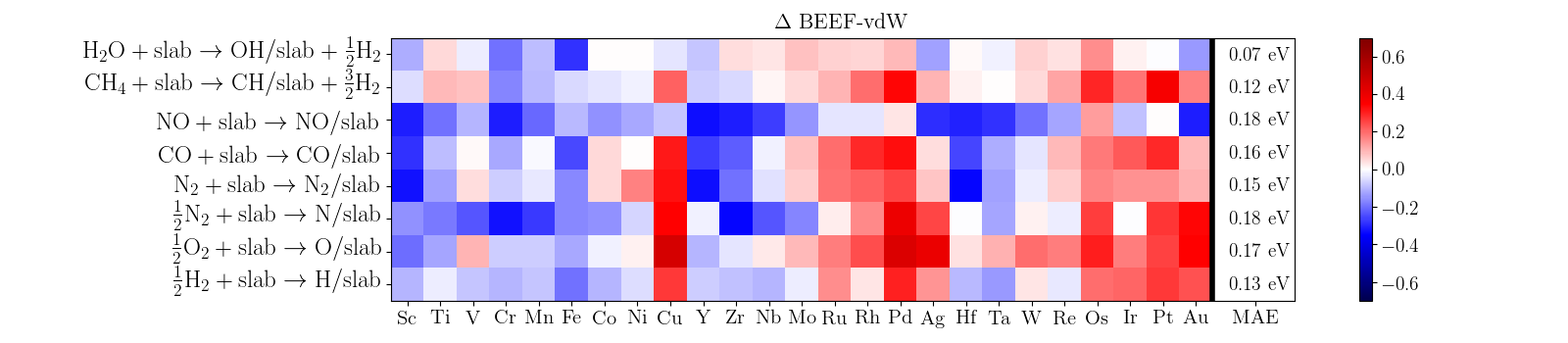}
\end{subfigure}
\begin{subfigure}{2.1\columnwidth}
\includegraphics[clip, trim=1.8cm 0.cm 3.5cm 0.cm, width=0.9\columnwidth]{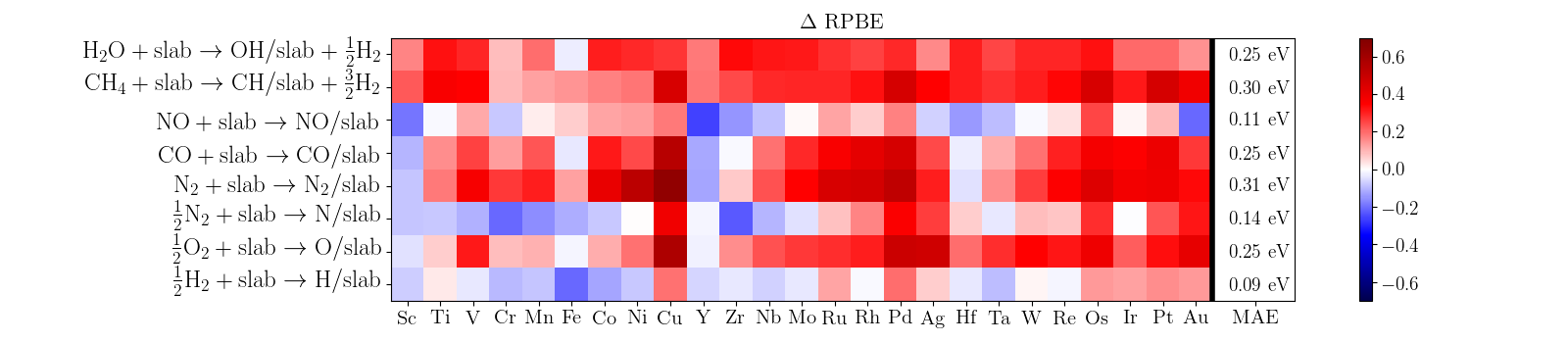}
\end{subfigure}
\hspace{0.2cm}
\caption{\label{fig:ads_RPA} Difference in adsorption energies calculated with PBE, RPBE and BEEF-vdW vs RPA. Blue (red) indicates overbinding (underbinding) relative to RPA.}
\end{figure*}

An overview of the deviation between the RPA and DFT reaction energies is given in Fig. \ref{fig:ads_RPA}. This shows how surface and reaction dependent the differences are. The subtitles should be understood as $\Delta \text{DFT} = E^{\text{DFT}}_\text{ads} - E^{\text{RPA}}_\text{ads}$. In this plot a blue color indicates an overbinding relative to RPA. The mean absolute error (MAE) for each reaction is show at the right of each column. Overall, we observe a rather large degree of variation between the various methods with energy differences ranging from around -0.6 eV to 0.6 eV. There is a general tendency of PBE to overestimate the binding while RPBE underestimates (relative to RPA). On the other hand the BEEF-vdW does not exhibit obvious systematic deviations from RPA although large absolute deviations of up to 0.4 eV are observed for some reactions. There are, however, some clear deviations from these general trends. For example, PBE underestimates the energies of reactions (1) and (2) on all surfaces except for Cr, Mn, Fe. In addition it underestimates most of the reaction energies on the noble metal surfaces. Similarly, RPBE overbinds, or underbinds less, on the early transition metals. 

Despite of the trends discussed above it is clear in general that the deviations between the DFT functionals and the RPA are both reaction and surface dependent. This indicates that the deviations have a non-trivial origin and cannot be gauged away by simple correction schemes such as tweaking the energies of the molecular reference states. 

\begin{figure*}[b]
\includegraphics[width=2.\columnwidth]{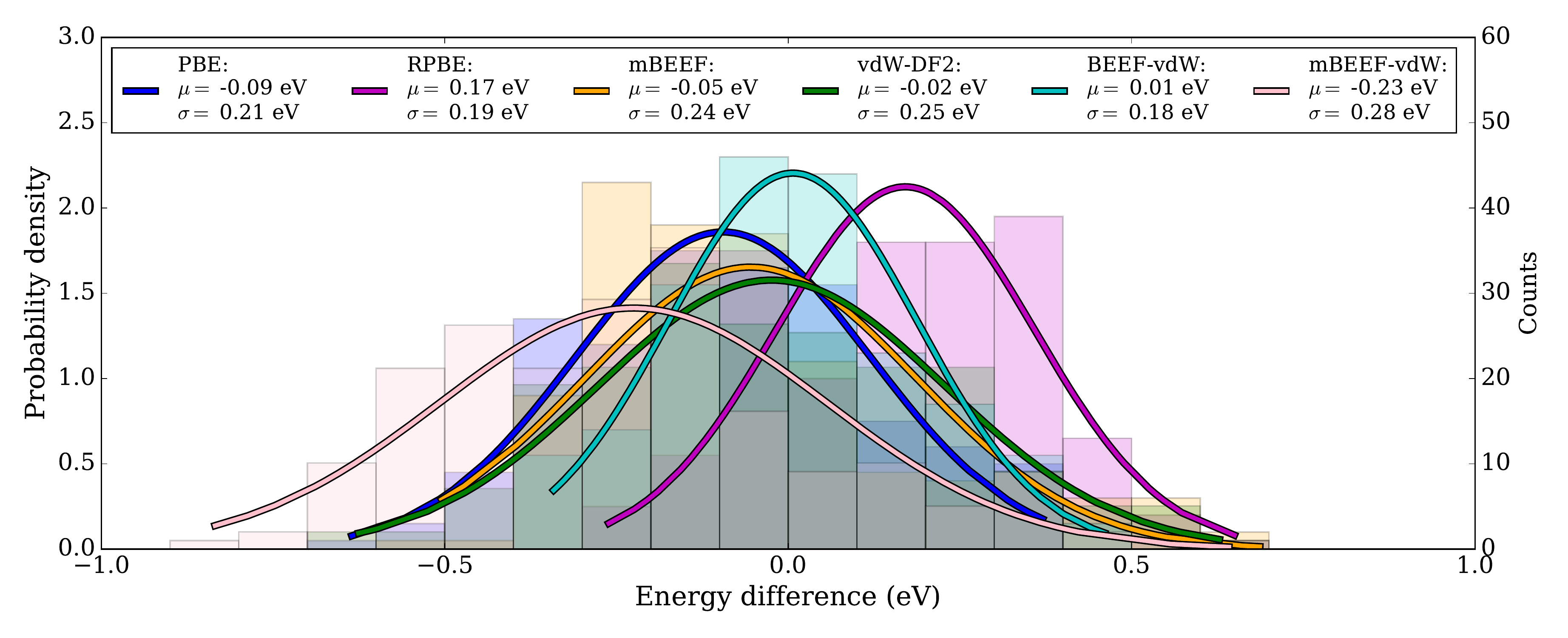}
\caption{\label{fig:distrib} Distribution of the deviations in adsorption energies between RPA and the DFT functionals employed.}
\end{figure*}

While the average deviation between RPA and the DFT functionals of around 0.2 eV, is rather modest, it should be kept in mind that this is an average over many reactions. For the individual reactions/surfaces the deviation can be as large as 0.6 eV. This is clear from Fig. \ref{fig:distrib} which shows the distributions of deviations between RPA and the DFT functionals. The mean of these distributions ($\mu$) correspond to the mean signed error (MSE). The standard deviation ($\sigma$) of the distributions are indicated.
LDA shows a significantly larger mean deviation of 0.72 eV with very systematic overbinding and is therefore not included in the figure. The fitted combined GGA/meta-GGA/vdW-DF functional mBEEF-vdW shows a mean of -0.23 eV and as such a systematic overbinding. The fitted BEEF-vdW functional is seen to perform the best, with a mean of 0.01 eV and a MAE of 0.14 eV.
Compared to the RPA reference, the PBE and RPBE perform equally well on average (MAE of 0.2 eV) but exhibit opposite systematic deviations. While the PBE tend to underbind (MSE of -0.09 eV) the RPBE shows a very systematic tendency to overbind (MSE 0.17 eV). 

\begin{figure*}
\includegraphics[width=1.65\columnwidth]{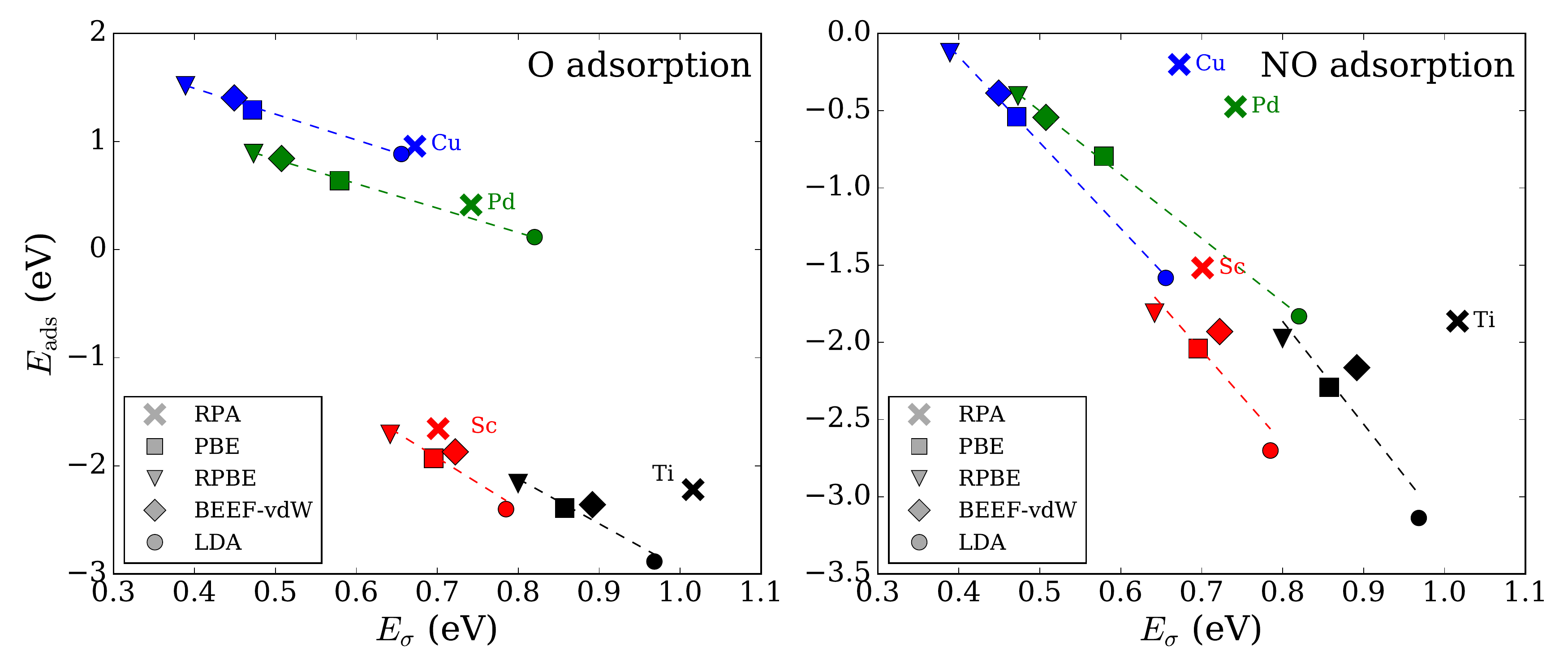}
\caption{\label{fig:DFTline_RPAout} Comparing adsorption and surface energies calculated with RPA (cross), PBE (square), RPBE (triangle), BEEF-vdW (diamond) and LDA (circle) for the adsorption of O and NO on four different surfaces.}
\end{figure*}

Fig. \ref{fig:DFTline_RPAout} shows the reaction energy versus surface energy for two different reactions on four different surfaces calculated with five different methods. As shown previously\cite{kresse_ads} for the specific case of CO on Pt(111), the DFT results fall roughly on a straight line supporting the statement that by tuning the xc-functional it is possible to change the adsorption and surface energies, but only simultaneously. In contrast, RPA is seen to deviate from this universal DFT line. The magnitude of the deviation is seen to be both surface and reaction dependent, again highlighting that RPA captures elements of the metal-adsorbate bonding mechanism missed by the DFT functionals. In some cases the BEEF-vdW results deviate slightly from the DFT line, suggesting that part of the additional physics captured by the RPA is related to van der Waals forces.

\begin{figure*}
\includegraphics[width=1.35\columnwidth]{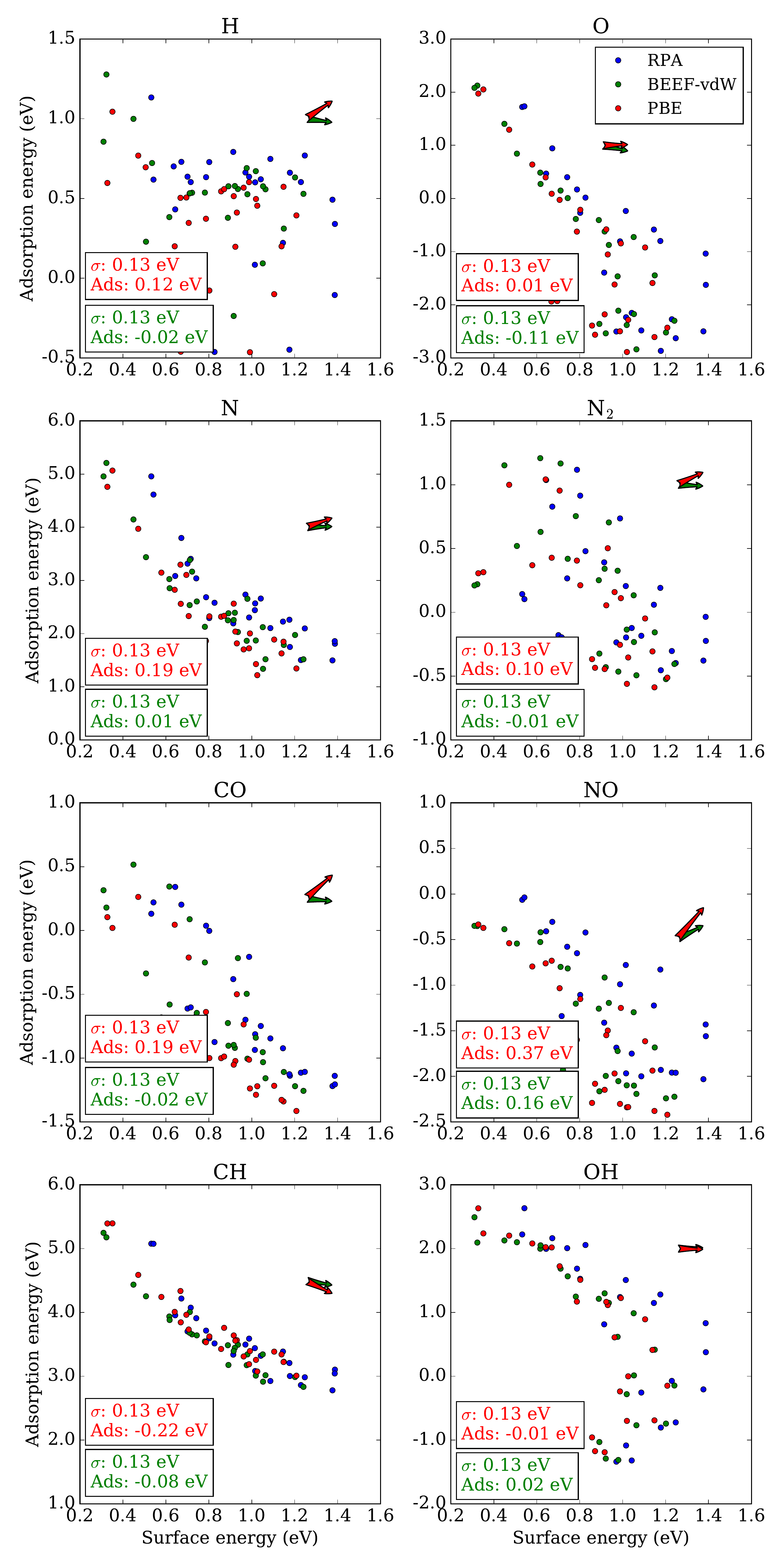}
\caption{\label{fig:ads_surf} The surface energy is plotted against the adsorption energy calculated with RPA (red), BEEF-vdW (green) and PBE (blue). Each dot represents a surface. The size and direction of the green (blue) arrow shows the difference between RPA and BEEF-vdW (PBE). The numbers in the boxes correspond to the length of the arrow in the x and y direction.}
\end{figure*}

Fig. \ref{fig:ads_surf} shows the adsorption energy versus surface energy for all the 8 reactions on all surfaces. The calculations have been performed using PBE (blue), BEEF-vdW (green) and RPA (red). The average difference between the DFT and RPA values is indicated by the arrows. The arrow thus indicates the mean signed error (MSE) which is a measure of the systematic deviation between the DFT and RPA values for each reaction. It is clear that both PBE and BEEF-vdW exhibit a systematic underestimation of the surface energy of 0.13 eV. 

\subsubsection{Scaling relations}
\begin{figure}
\includegraphics[width=\columnwidth]{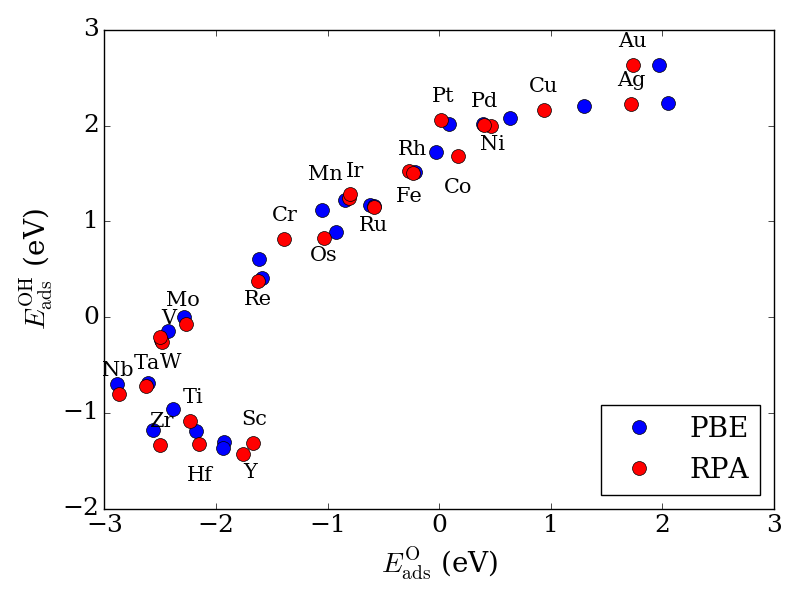}
\caption{\label{fig:scalingrel} Scaling relations between the adsorption energies of O and OH calculated with PBE and RPA. Deviations from the linear scaling relations are seen for both methods for the early and late transition metals.}
\end{figure}

As a final application we use the RPA data to explore the scaling relations between the adsorption energy of different adsorbates\cite{scalingrel}. The existence of different scaling relations have been established for several species on the basis of GGA calculations and have been exploited for various descriptor based approaches to catalyst design\cite{Norskov2009}. Fig. \ref{fig:scalingrel} shows the adsorption energy of OH versus the adsorption energy of O on a range of transition metal surfaces calculated with PBE and RPA. It is clear that even though the two methods deviate by around 0.2 eV on average, the scaling relations are seen to hold in both cases. This is in fact not surprising since the scaling relations are only fulfilled on a rather large energy scale on which the differences between PBE and RPA plays no role. It should be noted that significant deviations from linear scaling occur for both early and late transition metals. This happens because O and OH accept two and one electron from the metal, respectively. Consequently, transition metals with only one hole/electron in the $d$ band will bind the two species differently. This effect is expected to be coverage dependent. 

\section*{Conclusion}
In conclusion, we have performed a critical assessment of the performance of the RPA for chemisorption on transition metal surfaces. By studying 10 different reactions in the low coverage regime, the RPA was found to perform as well as the best GGA and van der Waals density functionals (RPBE and BEEF-vdW) for adsorption. The deviation between these methods and the experiments (mean absolute error around 0.2 eV) is comparable to or only slightly larger than the accuracy of the experiments themselves. For individual reactions rather large differences of up to 0.5 eV exist between the RPA and DFT adsorption energies. For surface energies, the RPA outperforms all the DFT functionals showing systematically better agreement with experiments.  
These results establishes the RPA as a universally accurate total energy method for surface science problems. 

The second part of the paper presented an RPA database of 200 adsorption energies for 8 different reactions on 25 different transition and noble metal surfaces.    To make this task tractable, these calculations were performed for a minimalistic model consisting of three metal layers and full coverage. It was, however, demonstrated that results from the minimal model were in qualitative agreement with more realistic models including more metal layers and lower coverage. In particular, the relative ordering of the binding energies obtained with different methods stays the same at high and low coverage. Some quantitative variation in binding energy versus coverage is observed. Of all methods, the RPA shows the smallest decrease in binding energy with coverage, which we ascribe to its incorporation of attractive dispersive interactions between the adsorbates.   

Our results unambiguously show that the RPA accounts for elements in the physics of the metal-adsorbate bonding and adsorbate-adsorbate interactions not captured by any of the investigated DFT functionals (GGA, meta-GGA, non-local vdW-DF). We believe that the comprehensive and systematically constructed data set of RPA adsorption energies will be useful for benchmarking and development of new and improved xc-functionals targeting the surface science community. 

\clearpage

\section{Method section}
\subsection{Computational implementation: The random phase approximation}
We employ the implementation of RPA in the electronic structure code Grid-based Projector-Augmented Wave method (GPAW)\cite{gpaw}, as described in greater detail in \cite{tolsrpa2}. In short, within RPA the xc energy contribution to the total ground state energy is split into an exact-exchange term plus RPA correlation. The exact exchange term can be written using plane waves as:
\begin{align}
E_x &= -\frac{1}{N_\mathbf{q} N_\mathbf{k}} \sum_{n,n'}\; \sum_{\mathbf{k},\mathbf{q}}^{ 1. \mathrm{BZ}} f_{n\mathbf{k}} \theta(\epsilon_{n\mathbf{k}} - \epsilon_{n'\mathbf{k}+\mathbf{q}}) \\
&\sum_\mathbf{G} v_\mathbf{G}(\mathbf{q}) |\braket{\psi_{n\mathbf{k}}|e^{-i(\mathbf{q}+\mathbf{G})\cdot \mathbf{r}}|\psi_{n'\mathbf{k}+\mathbf{q}}}|^2
\end{align}

where the number of plane waves is determined by a cutoff energy, $|\mathbf{q}+\mathbf{G}|^2/2 < E_{\mathrm{cut}}$.
The correlation energy contribution to the total energy is calculated from the non-interacting response function, $\chi^0(i\omega)$, by:
\begin{equation}
E_c^{\text{RPA}} = \int_0^\infty \frac{\mathrm{d}\omega}{2\pi} \text{Tr}\bigg[ \ln(1-\chi^0(i\omega)v) + \chi^0(i\omega)v\bigg]
\end{equation}

where $v$ is the Coulomb interaction and Tr is the trace. The response function and Coulomb interaction are evaluated in a plane wave basis and for periodic systems the trace involves a summation over \textbf{q}-points, which are determined from the Brillouin zone sampling. The frequency integration is carried out using a Gaussian quadrature. In a plane wave basis, the non-interacting response function is given as
\begin{align}
\chi_{\mathbf{G},\mathbf{G}'}^0&(\mathbf{q},i\omega) = \frac{1}{\Omega} \sum_\mathbf{k}^\text{BZ} \sum_{n,n'} \frac{f_{n\mathbf{k}} - f_{n'\mathbf{k}+\mathbf{q}}}{i\omega + \epsilon_{n\mathbf{k}} - \epsilon_{n'\mathbf{k}+\mathbf{q}}} \notag\\
&\times\braket{\psi_{n\mathbf{k}}|e^{-i(\mathbf{q}+\mathbf{G})\cdot \mathbf{r}}|\psi_{n'\mathbf{k}+\mathbf{q}}} \braket{\psi_{n'\mathbf{k}+\mathbf{q}}|e^{i(\mathbf{q}+\mathbf{G}')\cdot \mathbf{r}'}|\psi_{n\mathbf{k}}}
\end{align}
where $\Omega$ is the volume of the unit cell and $f_{n\mathbf{k}}$ occupation numbers. The sum over states is in principle infinite but in practice truncated at a finite number determined by the cutoff energy and the resulting total energy is then extrapolated using the usual $1/E_{cut}^{3/2}$ scheme
$$ E_c = E_c^\infty + \frac{A}{E_{cut}^{3/2}}$$
The rALDA calculations employ a renormalized adiabatic LDA exchange-only kernel which is described in great detail and applied to various systems in \cite{tols1,tols2,tols3}.
The adiabatic LDA kernel is given by 
$$ f_{xc}^\text{ALDA}[n](\mathbf{r},\mathbf{r}') = \delta(\mathbf{r}-\mathbf{r}') f_{xc}^\text{ALDA}[n]$$
where 
$$f_{xc}^\text{ALDA}[n] = \frac{d^2}{dn^2} \bigg( ne_{xc}^\text{HEG} \bigg) \bigg\vert_{n=n(\mathbf{r})},$$ 
The rALDA kernel is defined for the HEG by setting $f^{\text{rALDA}}_{xc}[n](q)=f^{\text{ALDA}}_{xc}[n]$ for $q<2k_F[n]$ and $-v(q)$ otherwise (this ensures continuity at $q=2k_F$). This results in a non-local kernel with the (almost) exact asymptotic $q\to \infty$ behaviour and without the divergences of the ALDA kernel~\cite{chris}.

The resulting response function is found by solving the Dyson equation
$$ \chi^\lambda(\omega) = \chi^0(\omega) + \chi^0(\omega)\bigg[\lambda v + f_{xc}^\lambda(\omega)\bigg]\chi^\lambda(\omega)$$
and the correlation energy is obtained through a numerical coupling constant integration
$$ E_c = -\int_0^1 d\lambda \int_0^\infty \frac{d\omega}{2\pi} \text{Tr} \bigg\lbrace v\bigg[ \chi^\lambda(i\omega) - \chi^0(i\omega)\bigg]\bigg\rbrace$$

The same $1/E_{cut}^{3/2}$ scheme is applied to the rALDA method.

\subsection{Computational details} \label{sec:compdetails}
\subsubsection{Experimental comparison}
The surfaces are represented by four layers with the bottom two fixed by the bulk PBE lattice constant found in the supplementary material of \cite{jess}, and the top two layers relaxed together with the adsorbate. $10\mathrm{\AA}$ of vacuum was used between neighbouring slabs and the reference is the isolated spin-polarized molecule in a $6 \mathrm{\AA} \times 6 \mathrm{\AA} \times 6 \mathrm{\AA}$ box. The RPA calculations were carried out using $8\times8\times1$ \textbf{k}-points and an identical q-point grid for Brillouin zone integration. 16 frequency points were used for the frequency integration and the extrapolation to infinite cutoff was done from calculations at 200, 250 and 300 eV. 
The EXX and RPA energies were evaluated on top of PBE eigenvalues and orbitals. 
The calculations are carried out using the plane wave version of the GPAW with the associated 0.9.11271 atomic setups.

\subsubsection{Adsorption database}
The surfaces were modeled using three layers with the bottom two layers fixed at the fcc PBE lattice constants from \url{www.materialsproject.org} and the position of the top layer relaxed. 
The position of the adsorbate was then found by carrying out an additional relaxation while keeping all three surface layers fixed. All relaxations were carried out with the BFGS algorithm using the PBE approximation to the xc-functional with a force convergence criteria of $0.05\,\mathrm{eV/\AA}$. The electron temperature was 0.01 eV and spin-polarized calculations were performed for calculations involving Fe, Ni or Co. $5\,\mathrm{\AA}$ of vacuum was added to either side of the adsorbate to avoid artificial interactions between neighboring layers following convergence tests at both the DFT and RPA level. The adsorption energies are relative to the molecule in its gas phase and the calculations for the isolated molecules were carried out in a $6\,\mathrm{\AA}\times6\,\mathrm{\AA}\times6\,\mathrm{\AA}$ box fully relaxing the geometry with the PBE functional.

\begin{figure}[h!]
\includegraphics[width=0.9\columnwidth]{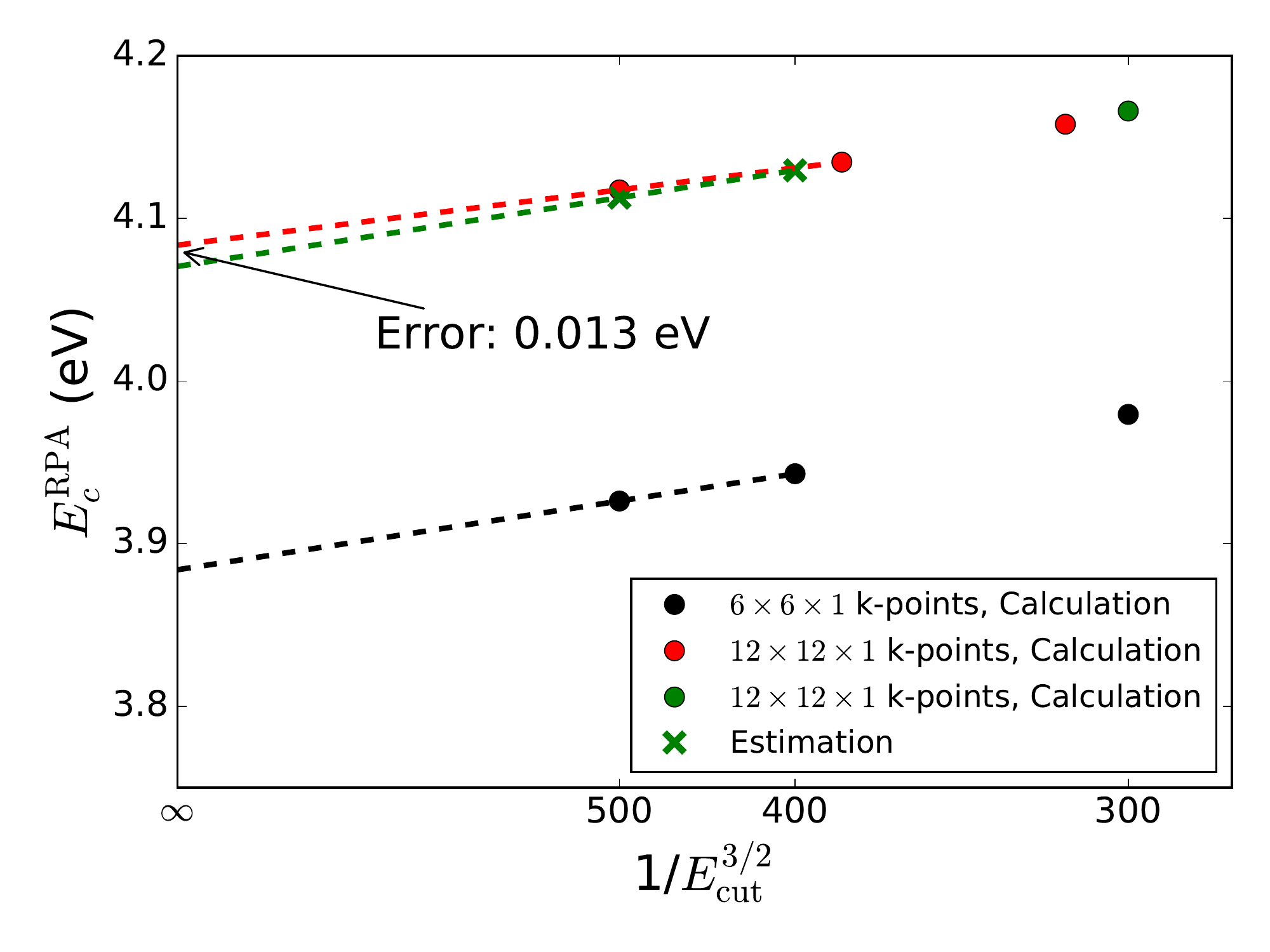}
\caption{\label{fig:extrap} Extrapolation of the correlation energy contribution to the total adsorption energy of O adsorbed on Mn. }
\end{figure}

The RPA calculations were carefully converged with respect to plane wave basis using the following extrapolation scheme:
In Fig. \ref{fig:extrap} the black dots are from a calculation with $6\times6\times1$ k-points (not enough to achieve convergence) but high cutoff energies (300, 400, 500 eV). The green circle is a calculation at a much denser k-point sampling of $12\times12\times1$ (converged). From these four circles, the two green crosses are predicted which allow for an extrapolation to infinite cutoff energy. The red dots represent actual calculations with both a dense \textbf{k}-point grid and high cutoff energies to test the extrapolation scheme. The error introduced by the extrapolation scheme for this particular system is seen to be 0.013 eV. The \textbf{k}-point grid of $12\times12\times1$ ensures that the exchange + correlation energy is converged to within 0.02 eV with respect to the \textbf{k}-point density.

\section{The Computational Materials Repository (CMR)}
The database is freely available via the Computational Materials Repository (CMR) at \url{https://cmr.fysik.dtu.dk/}. All structures, parameters and resulting energies can be found in the database file or browsed online. On the same website, it is explained in detail how to extract and use the data. 

\section{Acknowledgements}
The Center for Nanostructured Graphene is sponsored by the Danish National Research Foundation, Project DNRF58, the European Union’s Horizon 2020 research and innovation program under grant agreement no. 676580 with The Novel Materials Discovery (NOMAD) Laboratory, a European Center of Excellence; research grant 9455 from VILLUM FONDEN.

\section{Supporting Information}
In the supporting information we present tables of the adsorption and surface energies as found in the database. This material is available free of charge via the Internet at \url{http://pubs.acs.org}.
\clearpage

\bibliography{references}

\end{document}